\documentclass[a4paper]{jpconf}
\usepackage{amssymb}
\usepackage{graphicx}
\usepackage{epsfig}
\usepackage{slashed}

\begin{document}

\newcommand{\ppvq}{$pp \to W_H(Z_H) q_- + X$ }
\newcommand{\ppwq}{$pp \to W_H q_- + X$ }
\newcommand{\ppzq}{$pp \to Z_H q_- + X$ }
\newcommand{\qgwq}{$gq \to W_H q'_-$}
\newcommand{\qgvq}{$gq \to W_H(Z_H) q'_-$}
\newcommand{\qgwqx}{$gq \to W_H q'_- + X$}
\newcommand{\qgwqg}{$gq \to W_H q'_- + g$}
\newcommand{\qgwqq}{$gq \to W_H q'_- + q$}
\newcommand{\qqwqq}{$qq' \to W_H q''_- + q'$}
\newcommand{\ppqqwq}{$pp \to gg/q\bar{q} \to q'_- \bar{q}'_H \to W_H q'_- q''$ }
\newcommand{\qqwq}{$gg/q\bar{q} \to q_-' \bar{q}'_H \to W_H q'_- q''$}
\newcommand{\ppwqj}{$pp \to W_H q_- + j$}
\newcommand{\ppvqx}{$pp \to W_H(Z_H) q_- + X$}
\newcommand{\qgvqx}{$gq \to W_H(Z_H) q'_- + X$}
\newcommand{\qgvqg}{$gq \to W_H(Z_H) q'_- + g$}
\newcommand{\qgvqq}{$gq \to W_H(Z_H) q'_- + q$}
\newcommand{\qqvqq}{$qq' \to W_H(Z_H) q''_- + q'$}
\newcommand{\ppqqvq}{$pp \to gg/q\bar{q} \to q'_- \bar{q}'_H \to W_H(Z_H) q'_- q''$ }
\newcommand{\qqvq}{$gg/q\bar{q} \to q_-' \bar{q}'_H \to W_H(Z_H) q'_- q''$}
\newcommand{\ppvqj}{$pp \to W_H(Z_H) q_- + j$}
\newcommand{\etmiss}{\not\hskip-5truedd E_{T} }

\newcommand{\ppww}{$pp \to W_H^+ W_H^- + X$ }
\newcommand{\qqww}{$q\bar q \to W_H^+ W_H^-$}
\newcommand{\uuww}{$u\bar u \to W_H^+ W_H^-$}
\newcommand{\ggww}{$gg \to W_H^+ W_H^-~$}
\newcommand{\qqwwg}{$q\bar q \to W_H^+ W_H^- + g$}
\newcommand{\qgwwq}{$q(\bar q)g \to W_H^+ W_H^- + q(\bar q)$}

\title{Precise calculation for heavy gauge boson production in the LHT model }
\author{Guo Lei, Zhang Ren-You and Ma Wen-Gan}
\address{Department of Modern Physics, University of Science and Technology
of China (USTC), Hefei, Anhui 230026, P.R.China}

\ead{guolei@mail.ustc.edu.cn, zhangry@ustc.edu.cn, mawg@ustc.edu.cn}

\begin{abstract}
In the framework of the littlest Higgs model with $T$ parity, we
study the $W_H/Z_H+q_-$ and $W_H$-pair productions at the CERN Large 
Hadron Collider up
to the QCD next-to-leading order (NLO). The kinematic distributions
of final decay products and the theoretical dependence of the cross
section on the factorization/renormalization scale are analyzed. We
adopt the PROSPINO scheme in the QCD NLO calculations to avoid
double counting and keep the convergence of the perturbative QCD
description. By using the subtraction scheme,
the QCD NLO corrections  enhance the leading order
cross section with a K-factor in the range of $1.00 \sim 1.43$
for $W_H(Z_H) q_-$ production process, 
and in the range of $1.09 \sim 1.22$
for the $W_H$ pair production process.

\end{abstract}

\newcommand{\nb}{\nonumber}

\section{Introduction}
\par
Although the standard model (SM) \cite{s1,s2} provides a remarkably
successful description of high energy physics phenomena at the
energy scale up to $100~GeV$, it leaves a number of theoretical
problems unsolved. Many extended models are proposed to deal with
these problems such as grand unified theories \cite{model-1},
supersymmetric models \cite{model-2}, extra dimensions models
\cite{model-3}, left-right symmetric models \cite{model-5}, B-L (baryon
number minus lepton number)
extended SM models \cite{model-6}, little Higgs models
\cite{LittleHiggs} and many more. Each of these models has
motivation to solve one or more of the problems that the SM
encounters. Among them the little Higgs models deserve attention due
to their elegant solution to hierarchy problem and are proposed as
one kind of electroweak symmetry breaking (EWSB) models without
fine-tuning in which the Higgs boson is naturally light as a result
of nonlinearly realized symmetry \cite{Arkani}-\cite{LH7}. The
littlest Higgs (LH) model \cite{LH4}, an $SU(5)/SO(5)$ nonlinear
sigma model \cite{LH5}, is the most simplest version of little Higgs
models, in which a set of new heavy gauge bosons $(A_H,W_H,Z_H)$ and
a vector-like quark $(T)$ are introduced to cancel the quadratic
divergence contribution to Higgs boson mass from the SM gauge boson
loops and the top quark loop respectively. However, this model
predicts large corrections to electroweak precision observables and
the scale of the global symmetry breaking $f$, is constrained by
experimental data \cite{Limit-f}, which set severe constraints on
the new heavy particle masses and the model parameters. For
instance, recent experimental measurements on the decay processes of
$W_H^{\mp}\to l^{\mp}\stackrel{(-)}{\nu}$ and $Z_H\to l^{+}l^{-}$
provide the constraints of $M_{W_H}> 2.18~TeV$ and $M_{Z_H}>
1.83~TeV$ \cite{Wmass,Zmass}. These constraints would enforce the
symmetry breaking scale $f$, which characterizes the mass of new
particles, to be larger than $2.5~TeV$ and $3~TeV$ respectively.
Consequently, the cutoff scale $\Lambda \sim 4\pi f$ becomes so
large that calls for the fine-tuning between the electroweak scale
and the cutoff scale again.

\par
By introducing a discrete symmetry, the $T$ parity, the littlest
Higgs model with $T$ parity (LHT)
\cite{Low:2004xc}-\cite{Cheng:2003ju} offers a viable solution to
the naturalness problem of the SM, and also predicts a set of new
heavy fermions, gauge bosons as well as a candidate for dark matter.
In the LHT, all the SM particles are $T$-even and almost all the new
heavy particles are $T$-odd. Due to the different $T$ parity quantum
numbers, the SM gauge bosons cannot mix with the new gauge bosons in
the LHT. This would alleviate the constraints from the electroweak
precision tests and thus allows the scale $f$ to be significantly
lower than $1~TeV$ \cite{Hubisz:2005tx}. For instance, due to the
$T$ parity conservation, the processes $W_H^{\mp}\to
l^{\mp}\stackrel{(-)}{\nu}$ and $Z_H\to l^{+}l^{-}$ are forbidden,
and the only decay modes of these $T$-odd heavy gauge bosons are
$W_H\to A_H W$ and $Z_H\to A_H H$. In this case, the leptons are
produced from the decays of $W$ and $H$, but not from the heavy
gauge bosons directly. Therefore, these $T$-even gauge bosons escape
from the experimental constraints shown in Refs.\cite{Wmass,Zmass}.
Furthermore, as a lightest $T$-odd particle, the heavy photon $A_H$
cannot further decay into other particles, and would be a good
candidate for the dark matter \cite{Asano}. Since the CERN Large
Hadron Collider (LHC) has potential to detect the signals of new
gauge bosons and fermions, the phenomenology of the LHT would be
quite interesting and a number of phenomenological works has been
presented
\cite{Hubisz:2004ft,cpyuan:2006ph,sasha_pheno,Chen:2006ie}.

\par
In this paper, we present the QCD NLO corrections to the processes
\ppvq \cite{YanH} and \ppww \cite{DuSM}. 

\section{The related LHT theory }\label{theory}
\par
Before our calculations, we will briefly recapitulate the LHT theory
which is relevant to our work. The details of
the LHT can be found in
Refs.\cite{Low:2004xc,Hubisz:2004ft,Hubisz:2005tx,cpyuan:2006ph}.

\par
At some high scale $f$ the global symmetry $SU(5)$ is broken down to
$SO(5)$, leading to 14 massless Nambu-Goldstone bosons. Four of them
are manifested as the longitudinal modes of the heavy gauge bosons.
The other 10 decompose into a $T$-even $SU(2)$ doublet $h$, identified
as the SM Higgs field, and a complex $T$-odd $SU(2)$ triplet
$\Phi$, which obtains a mass of $m_{\Phi} = \sqrt{2}m_hf/v_{SM}$,
with $m_h$ and $v_{SM}$ being SM Higgs mass and the electroweak
symmetry break scale, respectively.

\par
The additional discrete symmetry, $T$-parity, is in analogy to the
$R$-parity in the minimal supersymmetric standard model (MSSM)
\cite{Low:2004xc,Hubisz:2004ft,Cheng:2003ju}. The $T$-parity
transformations for gauge sector are defined as the exchange between
the gauge bosons of the two $SU(2)\times U(1)$ groups, i.e., $W_1^a
\leftrightarrow W_2^a$ and $B_1 \leftrightarrow B_2$. Thus their
$T$-odd and $T$-even combinations can be obtained as
\begin{eqnarray}
W_H^a=\frac{1}{\sqrt{2}}(W_1^a-W_2^a),&
B_H=\frac{1}{\sqrt{2}}(B_1-B_2),& (T-odd), \nb\\
W_L^a=\frac{1}{\sqrt{2}}(W_1^a+W_2^a),&
B_L=\frac{1}{\sqrt{2}}(B_1+B_2),& (T-even).
\end{eqnarray}
The mass eigenstates of the gauge sector in the LHT are expressed as
\begin{eqnarray}
W_H^{\pm}=\frac{1}{\sqrt 2}(W_H^1\mp i W_H^2),& Z_H=s_H B_H+c_H
W^3_H,&
A_H=c_H B_H-s_H W^3_H , \nb \\
W_L^{\pm}=\frac{1}{\sqrt 2}(W_L^1\mp i W_L^2),& Z_L=-s_w B_L+c_w
W^3_L,& A_L=c_w B_L+s_w W^3_L,
\end{eqnarray}
where $s_{w}=\sin{\theta_{W}}$, $c_{w}=\cos{\theta_{W}}$,
$s_{H}=\sin{\theta_{H}}$, $c_{H}=\cos{\theta_{H}}$, $\theta_W$ is
the Weinberg angle, and the mixing angle $\theta_H$ at the ${\cal
O}(v^2/f^2)$ is expressed as
\begin{eqnarray}
\sin \theta_H \simeq \left[ \frac{5gg'}{4(5g^2-g'^2)}\frac{v_{SM}
^2}{f^2} \right].
\end{eqnarray}
Then the gauge sector consists of $T$-odd heavy new gauge bosons
$W_H^{\pm}$, $Z_H$, $A_H$ and $T$-even light gauge bosons identified
as SM gauge bosons, $W^{\pm}$, $Z^0$ and one massless photon. The
$T$ parity partner of the photon, $A_H$, is the lightest $T$-odd
particle, therefore, the candidate of dark matter in the LHT. The
masses of the $T$ parity partners of the photon, $Z^0$- and
$W^{\pm}$-boson are expressed as \cite{cpyuan:2006ph}
\begin{eqnarray} \label{m_v}
m_{W_H}\simeq m_{Z_H} \simeq
gf\left(1-\frac{1}{8}\frac{v_{SM}^2}{f^2}\right),& m_{A_H}\simeq
\frac{1}{\sqrt{5}}g'f\left(1-\frac{5}{8}\frac{v_{SM}^2}{f^2}\right),
\end{eqnarray}
where $v_{SM}= 246~GeV$. At the tree level the SM gauge boson masses
can be expressed as $m_W=\frac{gv_{SM}}{2}$ and
$m_Z=\frac{v_{SM}\sqrt{g^2+g^{\prime 2}}}{2}$.

\par
In the LHT, the fermion sector of the first two generations in the
SM is remained unchanged and the third generation of quarks is
modified. We introduce two fermion doublets $q_1$ and $q_2$ for each
fermion generation. The $T$ parity transformation to these fermion
doublets is defined as $q_1 \leftrightarrow  - q_2$. Therefore, the
$T$-odd and $T$-even combinations can be constructed as
$q_-=\frac{1}{\sqrt{2}}(q_1+q_2)$ and
$q_+=\frac{1}{\sqrt{2}}(q_1-q_2)$, where $q_+$ is the doublet for
the SM fermions and $q_-$ for their $T$-odd partners. We take the
Lagrangian suggested in
Refs.\cite{Low:2004xc,Hubisz:2004ft,Hubisz:2005tx} to generate the
masses of the $T$-odd fermion doublets,
\begin{eqnarray}
 -\kappa f (\bar{\Psi}_2 \xi \Psi_c
+\bar{\Psi}_1 \Sigma_0 \Omega \xi^\dagger \Omega\Psi_c)+{\rm h.c.},
\label{Lagrangian}
\end{eqnarray}
where $\Omega = diag(1,1,-1,1,1)$,
$\Psi_c=(q_c,\chi_c,\tilde{q}_c)^T$, and the $SU(5)$ multiplets
$\Psi_1$ and $\Psi_2$ are expressed as
\begin{equation}\label{Psi}
\begin{array}{ccc}
{\Psi}_1=\left( \begin{array}{c} q_1 \\ 0 \\ {\bf 0}_2
\end{array}\right) \,,& {\Psi}_2=\left(\begin{array}{c} {\bf 0}_2 \\
0 \\ q_2
\end{array}\right).
\end{array}
\end{equation}
The interaction Lagrangian in Eq.(\ref{Lagrangian}) can be proofed
to be invariant under $T$-parity, and $T$-odd quark doublet $q_-$
gets a Dirac mass with $\tilde{q}_c\equiv(id_{R_-},-iu_{R_-})^{\rm
T}$ from Eq.(\ref{Lagrangian}) expressed as \cite{cpyuan:2006ph}
\begin{eqnarray} \label{m_q}
m_{U_-}\simeq \sqrt{2}\kappa
f\left(1-\frac{1}{8}\frac{v_{SM}^2}{f^2}\right), &&
m_{D_-}=\sqrt{2}\kappa f,
\end{eqnarray}
where the lower indexes $U_-=u_-,c_-,t_-$ and $D_-=d_-,s_-,b_-$,
which represent the $T$-odd heavy partners of the SM quarks, and
$\kappa$ is the mass coefficient in Lagrangian of the quark sector.
As we know in the LHT $f > 500~GeV$ \cite{Hubisz}, it is evident
from Eq.(\ref{m_q}) that the $T$-odd up- and down-type heavy
partners have nearly equal masses.

\par
In order to avoid the large radiative correction to Higgs boson mass
induced by top-quark loop, the top sector must be additionally
modified. We introduce the following two multiplets,
\begin{equation}
\begin{array}{ccc}
{\cal Q}_1=\left( \begin{array}{c} q_1 \\ U_{L1} \\ {\bf 0}_2
\end{array}\right) \,,& {\cal Q}_2=\left(\begin{array}{c} {\bf 0}_2 \\
U_{L2} \\ q_2
\end{array}\right),
\end{array}
\end{equation}
where $U_{L1}$ and $U_{L2}$ are the singlet fields and the $q_1$ and
$q_2$ are the doublets. Under the $SU(5)$ and the $T$ parity
transformations, ${\cal Q}_1$ and ${\cal Q}_2$ behave themselves
same as $\Psi_1$ and $\Psi_2$.

\par
In addition to the $T$-even SM top quark right-handed $SU(2)$
singlet $u_R$, the LHT contains two $SU(2)$ singlet fermions
$U_{R1}$ and $U_{R2}$ of hypercharge 2/3, which transform under $T$
parity as
\begin{equation}
U_{R1}\leftrightarrow -U_{R2}.
\end{equation}
The T parity invariant Yukawa Lagrangian of the top sector can be
written as
\begin{eqnarray}\label{top-yukawa}
{\cal L}^{Y}_t &=& \frac{\lambda_1 f }{2\sqrt{2}}\epsilon_{ijk}
\epsilon_{xy} \big[ (\bar{{\cal Q}}_1)_i \Sigma_{jx} \Sigma_{ky}  -
(\bar{{\cal Q}}_2 \Sigma_0)_i \tilde{\Sigma}_{jx}
\tilde{\Sigma}_{ky} \big] u_R \nonumber \\ &&  +
\lambda_2 f (\bar{U}_{L1} U_{R1} + \bar{U}_{L2} U_{R2})+ {\rm
h.c.}~.
\end{eqnarray}
where $\tilde{\Sigma}=\Sigma_0\Omega \Sigma^\dagger \Omega \Sigma_0$
is the image of the $\Sigma$ field under $T$ parity, and $i,~j$ and
$k$ run over $1-3$ and $x$ and $y$ over $4-5$. The $T$ parity
eigenstates are constructed as
\begin{equation}\label{eigenstate}
q_\pm = \frac{1}{\sqrt{2}}(q_1 \mp q_2),~~~~ U_{L\pm} =
\frac{1}{\sqrt{2}}(U_{L1} \mp U_{L2}),~~~~ U_{R\pm} =
\frac{1}{\sqrt{2}}(U_{R1} \mp U_{R2}).
\end{equation}
The $T$-odd states $U_{L-}$ and $U_{R-}$ combine to form a Dirac
fermion $T_-$, and we obtain the mass of the $T_-$ quark from the
Lagrangian of Eq.(\ref{top-yukawa}) as
\begin{equation}\label{T-oddMass}
m_{T_-}=\lambda_2 f.
\end{equation}
The left-handed (right-handed) top quark $t$ is a linear combination
of $u_{L_+}$ and $U_{L_+}$ ($u_{R+}$ and $U_{R_+}$), and another
independent linear combination is a heavy $T$-even partner of the
top quark $T_+$:
\begin{eqnarray}
\left(
\begin{array}{c}
t_X\\
T_{+X}
\end{array}
\right)&=&
\left(
\begin{array}{cc}
c_X & -s_X\\
s_X & c_X
\end{array}
\right) \left(
\begin{array}{c}
u_{X_+}\\
U_{X_+}
\end{array}
\right),~~(X=L,R),
\end{eqnarray}
where the mixing matrix elements are approximately expressed as
\begin{eqnarray}
s_L = s_\alpha^2 \frac{v_{SM}}{f}+\cdots,~~ s_R = s_\alpha\left[
1-\frac{c_\alpha^2(c_\alpha^2-s_\alpha^2)}{2}\frac{v_{SM}^2}{f^2}+\cdots\right].
\label{s-LR}
\end{eqnarray}
There we define $s_\alpha=\lambda_1/\sqrt{\lambda_1^2+\lambda_2^2}$
and $c_\alpha=\lambda_2/\sqrt{\lambda_1^2+\lambda_2^2}$. The $t$ is
identified with the SM top and $T_+$ is its $T$-even heavy partner.
Then the masses of the top quark and $T$-even heavy top quark can be
obtained as
\begin{equation}\label{t&T-evenMass}
m_t\simeq \frac{ \lambda_1 \lambda_2
v_{SM}}{\sqrt{\lambda_1^2+\lambda_2^2}},~~~~m_{T_+} \simeq
f \sqrt{\lambda_1^2+\lambda_2^2}.
\end{equation}

The couplings of the T-odd $SU(2)$ doublet quarks and gauge bosons
to the T-even SM particles used in our calculations are listed in
Table \ref{tab1} \cite{Hubisz:2004ft,Blanke:2007ckm}, where
$(V_{Hu})_{ij}$ and $(V_{Hd})_{ij}$ are the matrix elements of the
CKM-like unitary mixing matrices $V_{Hu}$ and $V_{Hd}$,
respectively. The two mixing matrices satisfy
$V_{Hu}^{\dag}V_{Hd}=V_{CKM}$ \cite{Blanke:2007ckm}, therefore, they
cannot simultaneously be set to the identity. In the following
calculations we take $V_{Hu}$ to be a unit matrix, then we have
$V_{Hd}=V_{CKM}$.
\begin{table}[h]
\begin{center}
\begin{tabular}{|c|l||c|l|}
\hline
Interaction & ~~~~~~~~Feynman rule & Interaction & ~~~~~~~~~Feynman rule \\
\hline
&&& \\
$W_{H}^{+\mu} \bar{u}_-^i d^j$ &
$i\frac{g}{\sqrt{2}}(V_{Hd})_{ij}\gamma^\mu  P_L$ &
$W_{H}^{-\mu} \bar{d^i}_- u^j$ & $i\frac{g}{\sqrt{2}}(V_{Hu})_{ij}\gamma^\mu P_L$\\
&&& \\
$Z_{H}^{\mu} \bar{u}_-^i u^j$ & $i(\frac{g C_H}{2}-\frac{g'
S_H}{10})(V_{Hu})_{ij} \gamma^\mu P_L$ &
$Z_{H}^{\mu} \bar{d}_-^i d^j$ & $i(-\frac{g C_H}{2}-\frac{g' S_H}{10})(V_{Hd})_{ij} \gamma^\mu P_L$ \\
&&& \\
$\bar{q}_{-}^{\alpha} q_{-}^{\beta} G^{a}_{\mu}$ & $ig_s (T^a)_{\alpha\beta}\gamma^{\mu}$ && \\
&&& \\
\hline
\end{tabular}
\caption{\label{tab1} The related LHT Feynman rules used in our
calculations, where $q_-=u_-,d_-,c_-,s_-,t_-,b_-$, $i$ and $j$ are the
generation indices and $C_H^2=1-S_H^2$.}
\end{center}
\end{table}

\section{Renormalization and PROSPINO scheme}

The strong coupling constant, the masses and wave functions of the
relevant colored particles in the LHT are renormalized to remove the
UV divergences of the virtual corrections.
In our calculations,
the following renormalization constants are introduced:
\begin{eqnarray}
\label{defination of renormalization constants}
\psi_{q(q_-)}^{0,L,R} &=& \left( 1 + \frac{1}{2} \delta
Z_{q(q_-)}^{L,R} \right) \psi_{q(q_-)}^{L,R},~~~~~
m^{0}_{q_-} = m_{q_-} + \delta m_{q_-},~~ \nb \\
G_{\mu}^0
&=&
\left(
1 + \frac{1}{2} \delta Z_g
\right)
G_{\mu},~~~~~~~~~~~~~~
g_s^0 = g_s + \delta g_s,
\end{eqnarray}
where $g_s$ denotes the strong coupling constant, $m_{q_-}$ is the
T-odd quark mass, $\psi_{q(q_-)}^{L,R}$ and $G_{\mu}$ denote the
fields of the SM quark, T-odd heavy quark and gluon, respectively.
The masses and wave functions of the colored fields are renormalized
by adopting the on-shell scheme, then the relevant renormalization
constants are expressed as
\begin{eqnarray} \label{CT-q}
\delta Z_{q}^{L,R}
& \equiv & \delta Z_{q}
=
-\frac{\alpha_s(\mu_r)}{3 \pi}
\Big[
\Delta_{UV}-\Delta_{IR}
\Big],  \\
\delta Z_{q_-}^{L,R}
& \equiv & \delta Z_{q_-}
=
-\frac{\alpha_s(\mu_r)}{3 \pi}
\left[
\Delta_{UV}+2\Delta_{IR}+4+3\ln\left(\frac{\mu_r^2}{m_{q_-}^2}\right)
\right],  \\
\frac{\delta m_{q_-}}{m_{q_-}} &=& - \frac{\alpha_s(\mu_r)}{3 \pi}
\left\{ 3\left[ \Delta_{UV} + \ln\left( \frac{\mu_r^2}{m_{q_-}^2}
\right) \right]+4 \right \},  \\
\label{CT-Zg} \delta Z_g &=& - \frac{\alpha_s(\mu_r)}{2
\pi}\left\{\frac{3}{2}\Delta_{UV}+\frac{5}{6}\Delta_{IR}+
\frac{1}{3}\ln\left(\frac{\mu_r^2}{m_t^2}\right)
+\frac{1}{3}\sum\limits_{T=T_+}^{T_-}\ln\left(\frac{\mu_r^2}{m_{T}^2}\right)
+\frac{1}{3}\sum\limits_{q_-}\ln\frac{\mu_{r}^2} {m_{q_-}^2}\right\}, \nb \\
&&~~~~~~~~~~~~~~~~~~~~~~~~~~~~~~~~~~~~~~~~~~~~~~~~~~~~~~~~~
(q_-=u_-,d_-,c_-,s_-,t_-,b_-),
\end{eqnarray}
where $\Delta_{UV}=1/\epsilon_{UV} -\gamma_E +\ln(4\pi)$ and
$\Delta_{IR}=1/\epsilon_{IR} -\gamma_E +\ln(4\pi)$.

\par
For the renormalization of the strong coupling constant $g_{s}$, we
adopt the $\overline{MS}$ scheme at the renormalization scale
$\mu_{r}$, except that the divergences associated with the massive
top-quark, T-odd $SU(2)$ doublet quarks ($u_-, d_-, c_-, s_-, t_-,
b_-$) and $T_{\pm}$ loops are subtracted at zero momentum \cite{gs}.
Then the renormalization constant of the strong coupling constant
can be obtained as
\begin{eqnarray} \label{CT-g}
\frac{\delta g_s}{g_s}&=& -\frac{\alpha_s(\mu_r)}{4\pi}
\left[\frac{3}{2}\Delta_{UV}+\frac{1}{3}\ln\frac{m_{t}^2}
{\mu_{r}^2}+\frac{1}{3}\sum\limits_{T=T_+}^{T_-}\ln\frac{m_{T}^2}
{\mu_{r}^2}+ \frac{1}{3}\sum\limits_{q_-}\ln\frac{m_{q_-}^2}
{\mu_{r}^2}\right],  \nb \\
&& ~~~~~~~~~~~~~~~~~~~~~~~~~~~~~~~~~~~~~
~~~~~~~(q_-=u_-,d_-,c_-,s_-,t_-,b_-).
\end{eqnarray}

\par
In the calculation, we can find there are additional on-shell
T-odd quark resonance in the real light-quark emissions. We
adopt the PROSPINO scheme \cite{PROSPINO-ref,on-shell subtraction}
to remove them. The PROSPINO scheme is defined as a replacement
of the Breit-Wigner propagator \cite{on-shell subtraction}
\begin{eqnarray}
 \frac{|{\cal M}|^2( s_{V_H q} )}{( s_{V_H q} - m_{q_-}^2 )^2
 + m_{q_-}^2 \Gamma_{q_-}^2}
 & \to &
 \frac{|{\cal M}|^2( s_{V_H q} )}{( s_{V_H q} - m_{q_-}^2 )^2
 + m_{q_-}^2 \Gamma_{q_-}^2} \\
 &&-
 \frac{|{\cal M}|^2( m_{q_-}^2 )}{( s_{V_H q} - m_{q_-}^2 )^2
 + m_{q_-}^2 \Gamma_{q_-}^2}
 \Theta( \hat{s} - 4 m_{q_-}^2 )
 \Theta( m_{q_-} - m_{V_H} ), \nonumber
\end{eqnarray}
where $s_{V_H q}$ is the squared momentum flowing through the
intermediate $q_-$ propagator.

\section{Numerical results of \ppvq process}
Due to the additional T-odd quark resonance in light-quark emission
subprocesses, we apply three schemes in considering the QCD NLO
corrections in this work. In scheme (I) (denoted as ``QCD NLO I'') we include all
light-quark emission contributions in the QCD NLO corrections.
In scheme (II) (denoted as ``QCD NLO II'') we exclude the
contributions of the partonic processes of light-quark emission.
The PROSPINO scheme for light-quark emission is used in 
scheme (III).

\par
In the study of the dependence of the QCD NLO corrected cross
section on the factorization and renormalization scales, we set the
two unphysical scales equal to a common value ($\mu_f = \mu_r =
\mu$) and do not vary them in an independent way for simplicity.

\par
We take one-loop and two-loop running $\alpha_{s}$ in the LO and QCD
NLO calculations, respectively \cite{databook}. The central value of
the factorization/renormalization scale $\mu$ is chosen as
$\mu_0=(m_{W_H}+m_{d_-})/2$. We adopt the CTEQ6L1 and CTEQ6M parton
densities with five flavors in the LO and NLO calculations,
respectively \cite{cteq}. The strong coupling constant
$\alpha_s(\mu)$ is determined by the QCD parameter $\Lambda_5^{LO} =
165~MeV$ for the CTEQ6L1 at the LO and $\Lambda_5^{\overline{MS}} =
226~MeV$ for the CTEQ6M at the NLO \cite{databook}. We ignore the
masses of $u$-, $d$-, $c$-, $s$-, $b$-quarks, and take
$\alpha_{ew}(m_Z^2)^{-1}|_{\overline{MS}}=127.925$,
$m_W=80.399~GeV$, $m_Z=91.1876~GeV$, $m_t=171.2~GeV$ and
$\sin^2\theta_W=1-\left(\frac{m_W}{m_Z}\right)^2=0.222646$.

\par
The colliding energy in the proton-proton center-of-mass system is
taken as $\sqrt s=7~TeV$ for the early LHC and $\sqrt s=14~TeV$ for
the later running at the LHC. The Cabibbo-Kobayashi-Maskawa (CKM)
matrix elements are taken as
\begin{eqnarray}\label{CKM}
 V_{CKM} &=& \left(
\begin{array}{ccc}
    V_{ud} \ &  V_{us} \ &  V_{ub} \\
    V_{cd} \ &  V_{cs} \ &  V_{cb} \\
    V_{td} \ &  V_{ts} \ &  V_{tb} \\
\end{array}
    \right)=\left(
\begin{array}{ccc}
     0.97418 \ &  0.22577 \ &  0 \\
    -0.22577 \ &  0.97418 \ &  0 \\
       0 \ &  0 \ &  1 \\
\end{array}  \right).
\end{eqnarray}

\subsection{Dependence on factorization/renormalization scale \label{sub-0} }
\par
\begin{figure}[htbp]
\begin{center}
\includegraphics[width=0.32\textwidth]{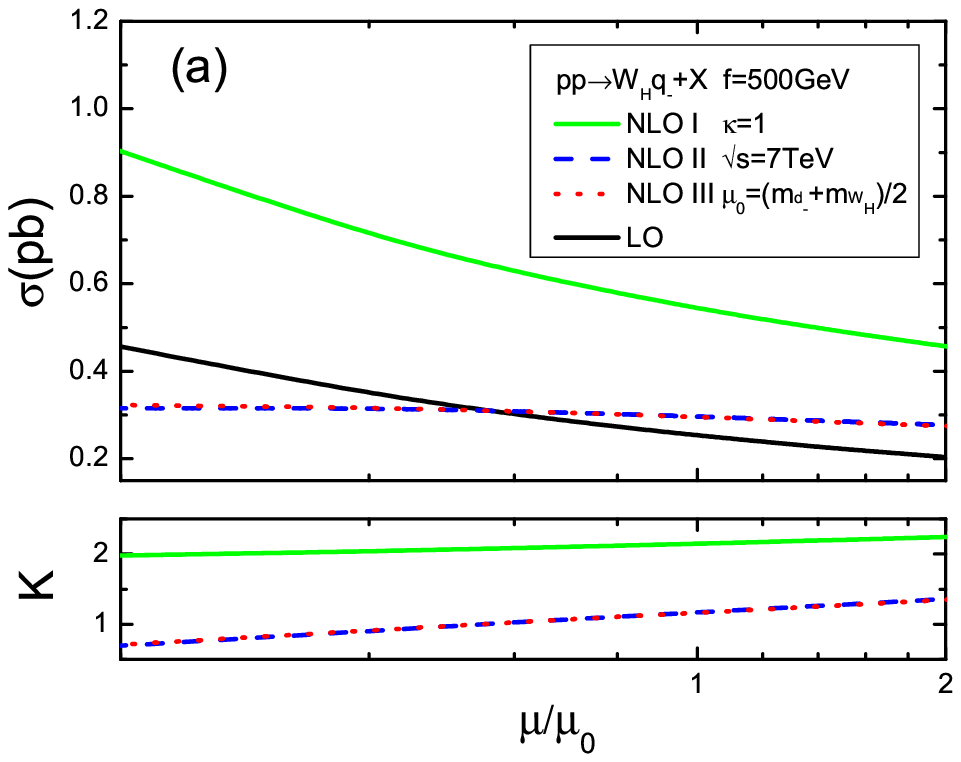}
\includegraphics[width=0.32\textwidth]{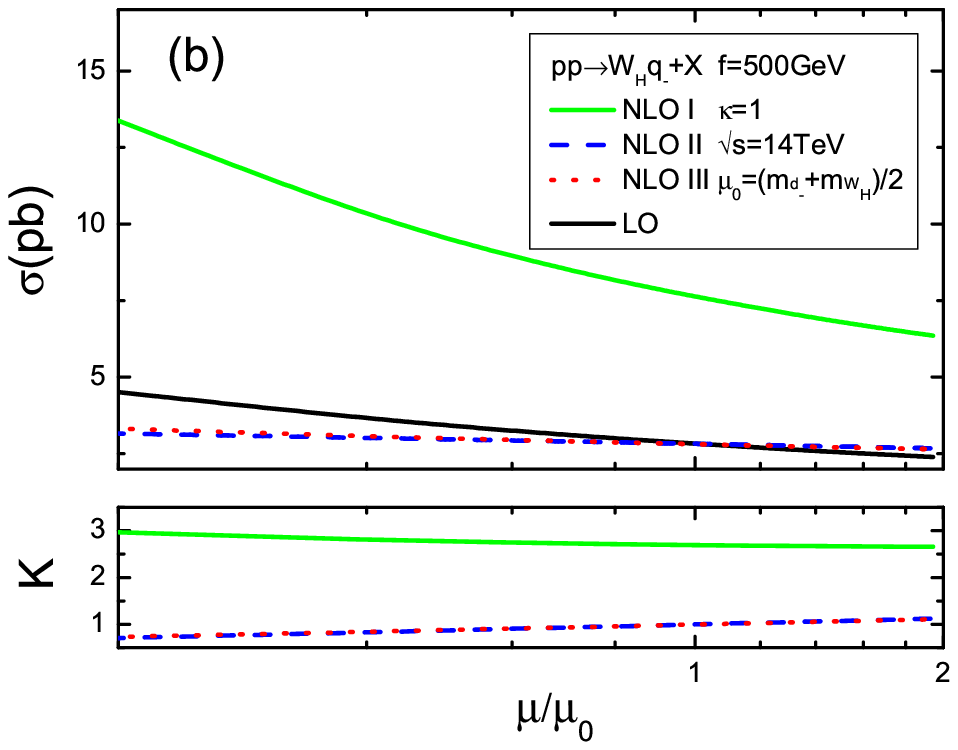}
\includegraphics[width=0.32\textwidth]{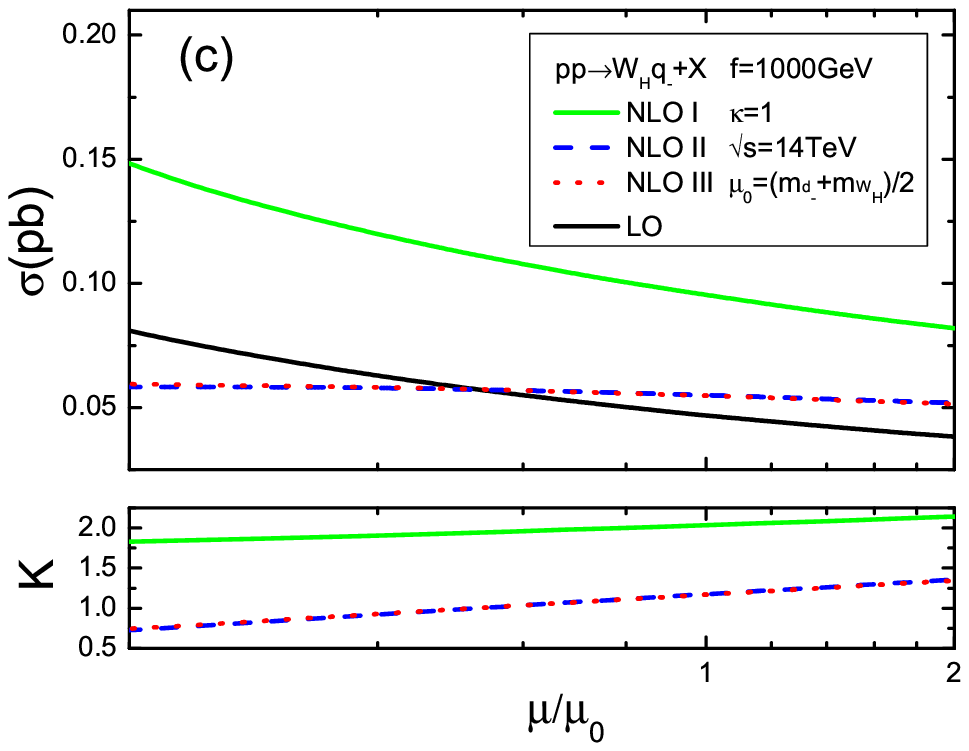}`
\caption{\label{figa5} The dependence of the cross sections and the
corresponding K-factors for the \ppwq
process on the factorization/renormalization scale $\mu$ at the LHC.
(a) $f=500~GeV$, $\kappa=1$ and $\sqrt{s}=7~TeV$. (b) $f=500~GeV$,
$\kappa=1$ and $\sqrt{s}=14~TeV$. (c) $f=1~TeV$, $\kappa=1$ and
$\sqrt{s}=14~TeV$.  }
\end{center}
\end{figure}
\begin{figure}[htbp]
\begin{center}
\includegraphics[width=0.32\textwidth]{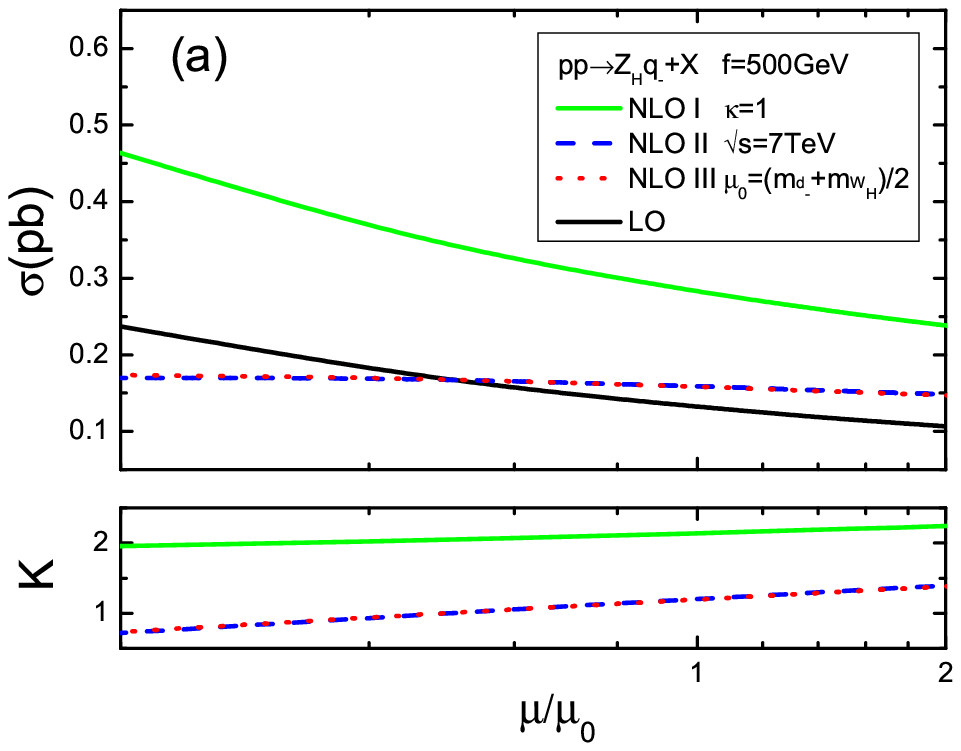}
\includegraphics[width=0.32\textwidth]{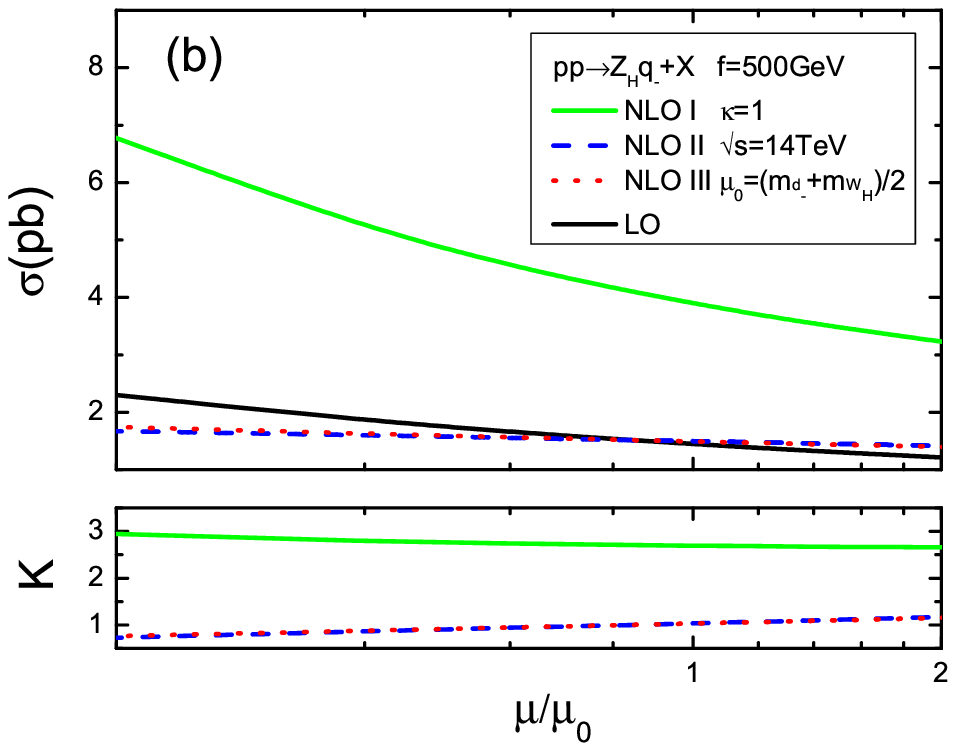}
\includegraphics[width=0.32\textwidth]{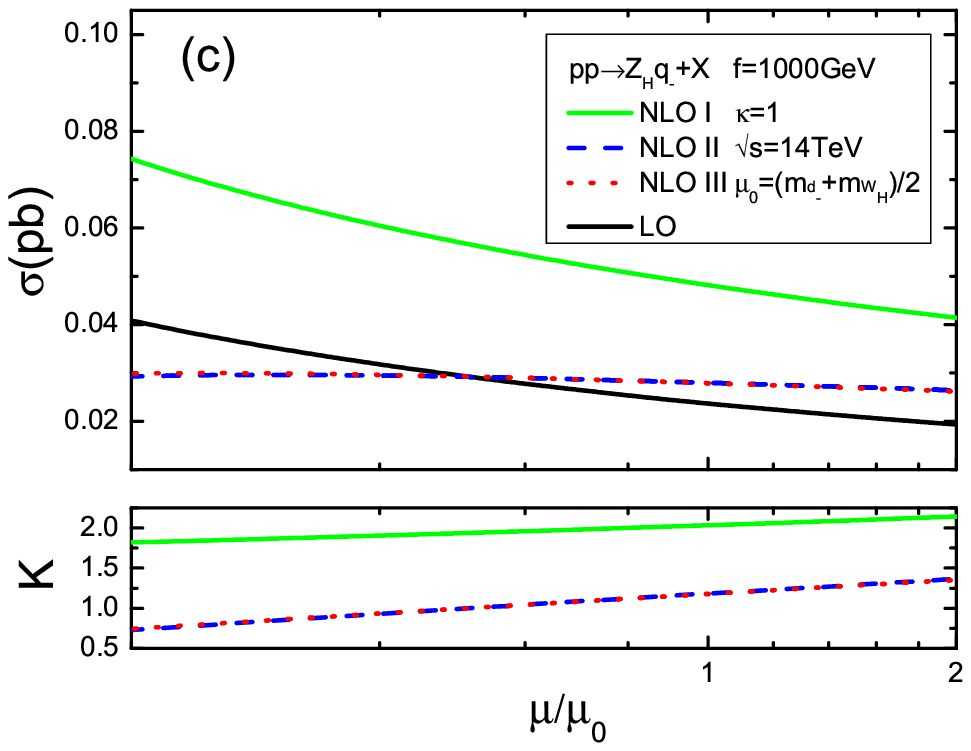}
\caption{\label{figa6} The dependence of the cross sections and the
corresponding K-factors for the \ppzq
process on the factorization/renormalization scale $\mu$ at the LHC.
(a) $f=500~GeV$, $\kappa=1$ and $\sqrt{s}=7~TeV$. (b) $f=500~GeV$,
$\kappa=1$ and $\sqrt{s}=14~TeV$. (c) $f=1~TeV$, $\kappa=1$ and
$\sqrt{s}=14~TeV$.  }
\end{center}
\end{figure}

In Figs.\ref{figa5}(a,b,c) and Figs.\ref{figa6}(a,b,c) we present the
LO, QCD NLO corrected cross sections and the corresponding K-factors
for the \ppwq and \ppzq processes as the functions of the
factorization/renormalization scale at the LHC with $\sqrt{s}=7~TeV$
and $14~TeV$, respectively. In Figs.\ref{figa5}(a,b) and
Figs.\ref{figa6}(a,b) the LHT input parameters are taken as
$f=500~GeV$ and $\kappa=1$, while in Fig.\ref{figa5}(c) and
Fig.\ref{figa6}(c) we take $f=1~TeV$ and $\kappa=1$.
In these figures the curves labeled by "NLO I", "NLO II" and "NLO
III" are for the QCD NLO corrected cross sections using the (I),
(II) and (III) schemes, respectively. The figures show that by using
the (II) and (III) subtraction schemes we can get almost the same
and moderate QCD NLO corrections to the production rate with a
strongly reduced factorization/renormalization scale uncertainty in
the plotted range of $\mu$, while the QCD NLO corrections using the
scheme (I) do not obviously improve the scale dependence of the LO
cross section and destroy the perturbative convergence in some range
of $\mu$. In the following analysis we set the
factorization/renormalization scale $\mu$ as its central value
$\mu_0=(m_{W_H}+m_{d_-})/2$.

\subsection{Dependence on LHT parameters \label{sub-1} }
\par
\begin{figure}[htbp]
\begin{center}
\includegraphics[width=0.32\textwidth]{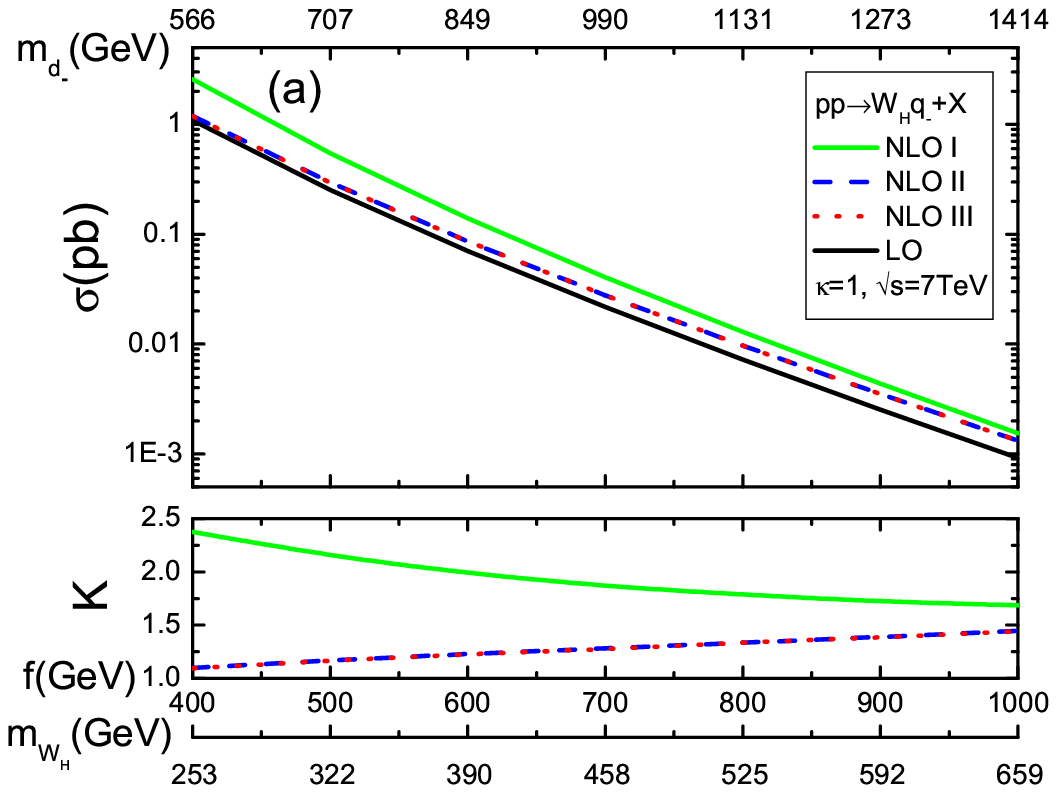}
\includegraphics[width=0.32\textwidth]{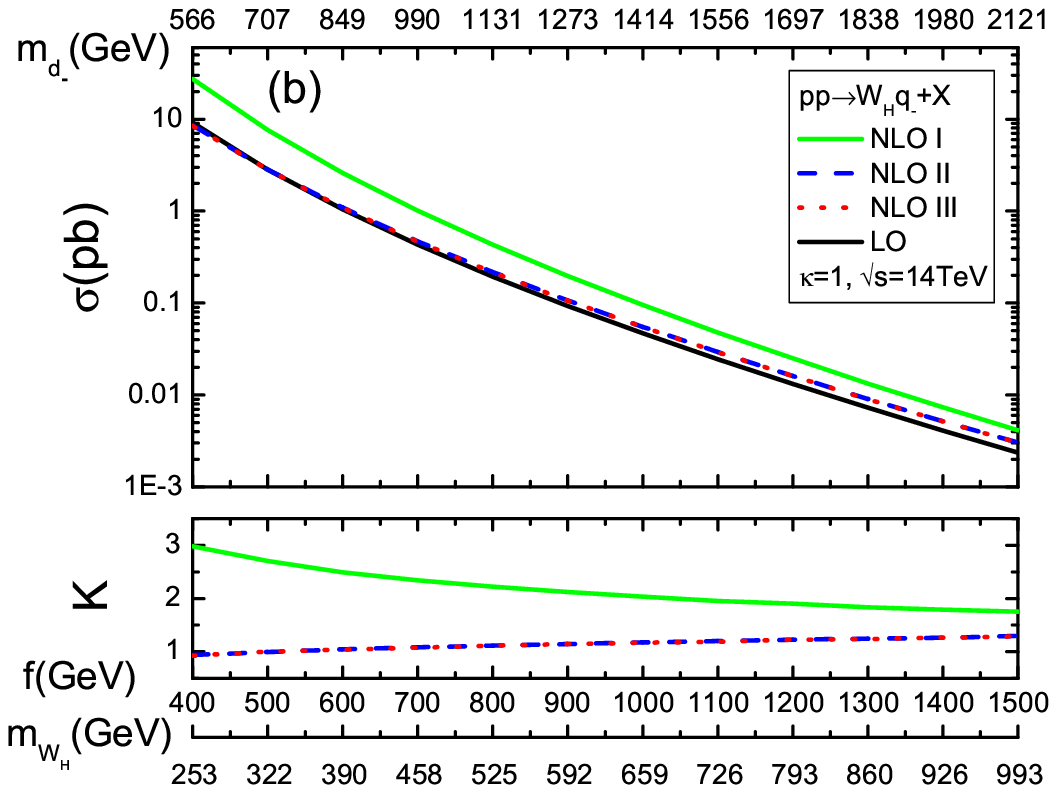}
\includegraphics[width=0.32\textwidth]{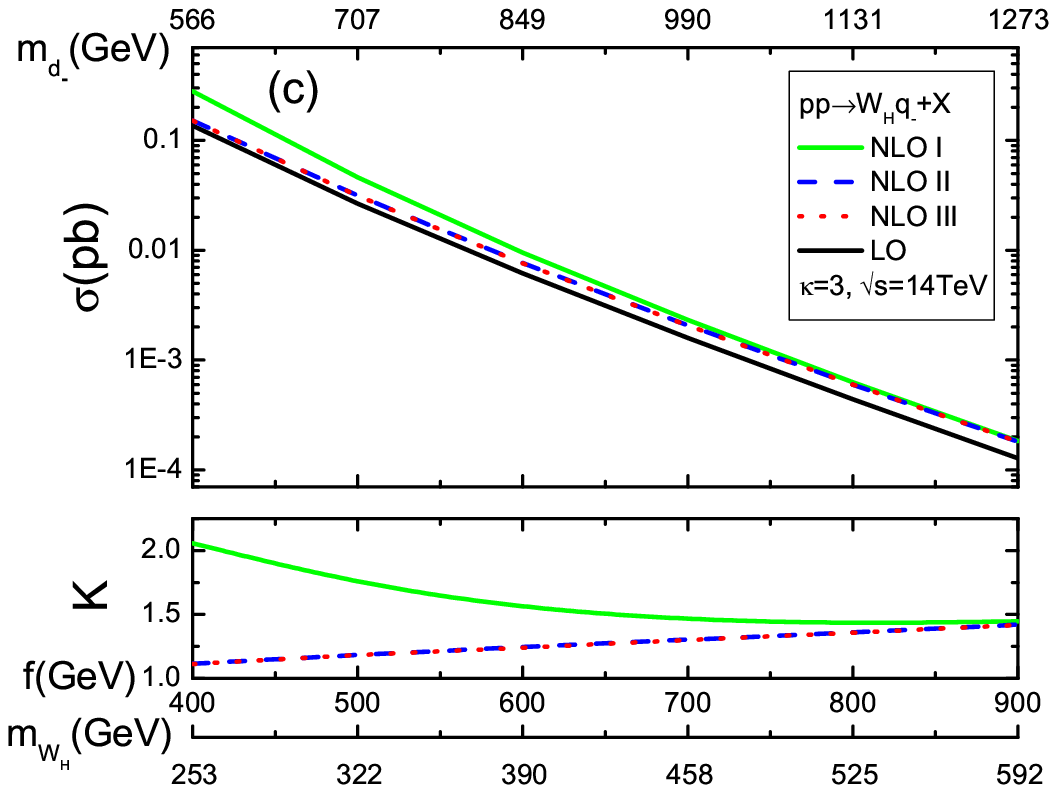}
\caption{\label{figa7} The cross sections and the corresponding
K-factors for the \ppwq process as the functions of the LHT
parameter $f$ at the LHC. The corresponding $m_{W_H}$ and $m_{d_-}$
values are also scaled on the x-axis. (a) $\kappa=1$ and
$\sqrt{s}=7~TeV$. (b) $\kappa=1$ and $\sqrt{s}=14~TeV$. (c)
$\kappa=3$ and $\sqrt{s}=14~TeV$. }
\end{center}
\end{figure}
\begin{figure}[htbp]
\begin{center}
\includegraphics[width=0.32\textwidth]{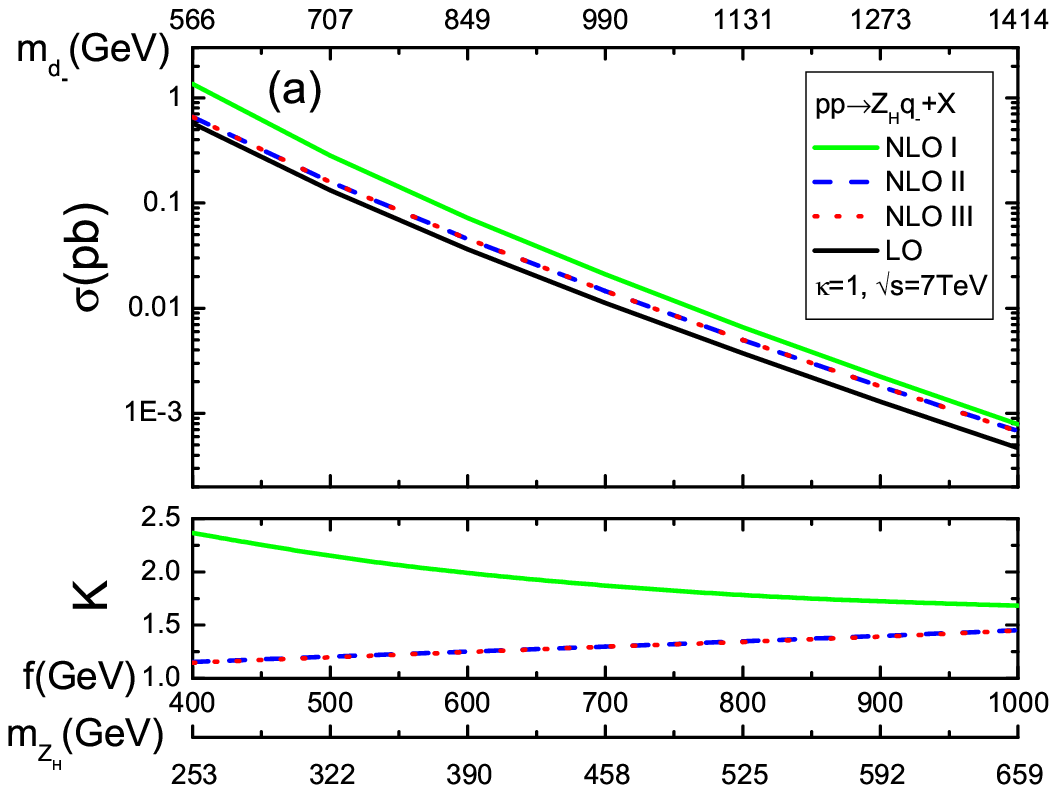}
\includegraphics[width=0.32\textwidth]{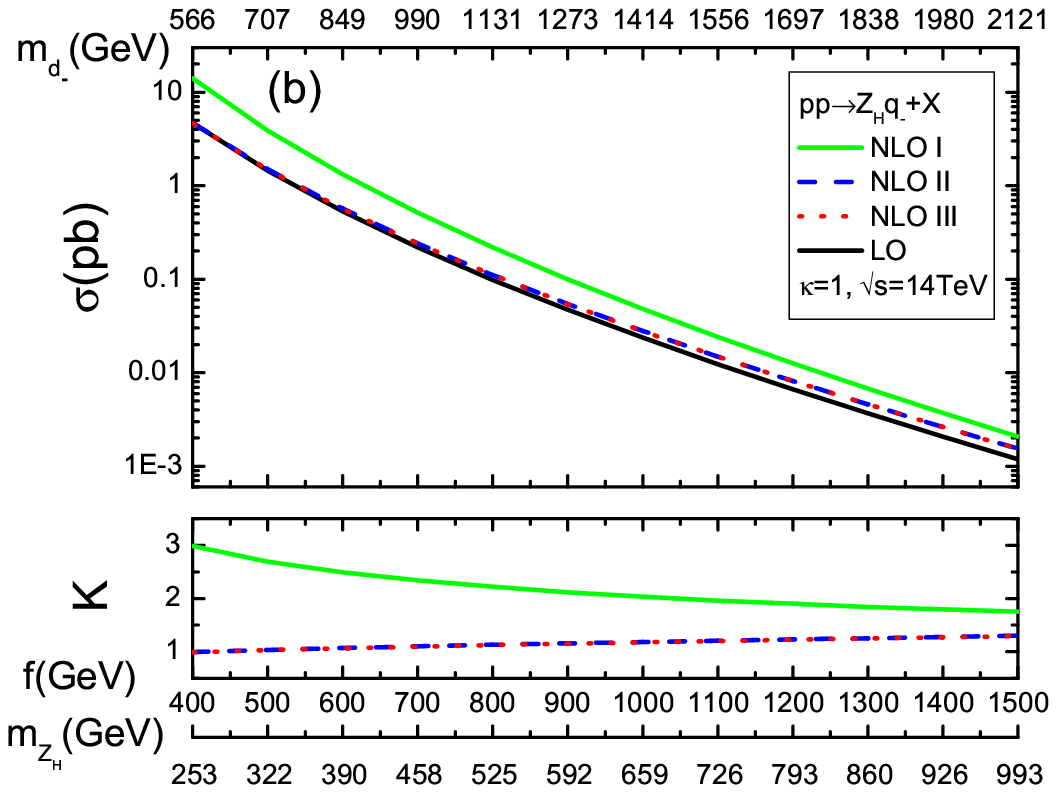}
\includegraphics[width=0.32\textwidth]{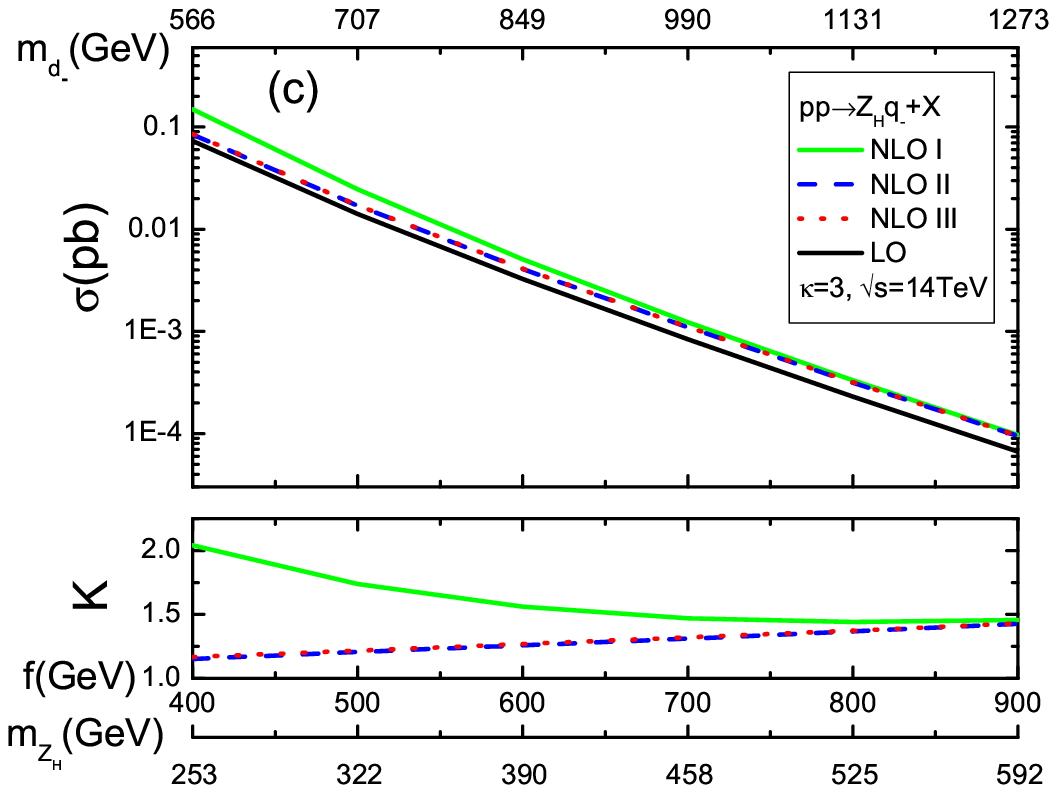}
\caption{\label{figa8} The cross sections and the corresponding
K-factors for the \ppzq process as the functions of the LHT
parameter $f$ at the LHC. The corresponding $m_{W_H}$ and $m_{d_-}$
values are also scaled on the x-axis. (a) $\kappa=1$ and
$\sqrt{s}=7~TeV$. (b) $\kappa=1$ and $\sqrt{s}=14~TeV$. (c)
$\kappa=3$ and $\sqrt{s}=14~TeV$. }
\end{center}
\end{figure}

We depict the LO, QCD NLO corrected cross sections and the
corresponding K-factors for the \ppwq and \ppzq processes as the
functions of $f$, the $SU(5)$ global symmetry breaking scale of the
LHT, at the LHC with $\sqrt{s}=7~TeV$ and $14~TeV$ in
Figs.\ref{figa7}(a,b,c) and Figs.\ref{figa8}(a,b,c), respectively. In
Figs.\ref{figa7}(a,b) and Figs.\ref{figa8}(a,b) the parameter $\kappa$
is set to be $1$, while in Fig.\ref{figa7}(c) and Fig.\ref{figa8}(c)
we take $\kappa=3$. The curves labeled by "NLO I", "NLO II" and "NLO
III" are for the QCD NLO corrected cross sections using the (I),
(II) and (III) schemes, respectively. One can conclude from these
figures that the cross section for the $pp \rightarrow W_H(Z_H) q_-
+ X$ process decreases quickly with the increment of $f$, because
the two final T-odd particles become heavier with the increment of
$f$. However, in the plotted range of $f$ we could have observable
production rates for the $p p \to W_H q_- + X$ and $p p \to Z_H q_-
+ X$ processes, especially when $\kappa=1$.

\section{Numerical results of \ppww process}

We take $\alpha_{ew}(m_Z^2)^{-1}=127.916$, $m_W=80.399~GeV$,
$m_Z=91.1876~GeV$,
$\sin^2\theta_W=1-\left(\frac{m_W}{m_Z}\right)^2=0.2226$ and
$m_t=171.2~GeV$ \cite{databook}. The masses of all the SM leptons
and quarks except top quark are neglected. The center-of-mass
energies $\sqrt{s}$ of proton-proton collision are taken to be
$14~TeV$ and $8~TeV$ for the future and early LHC, separately. We
set the factorization and renormalization scale to be equal ($\mu_r
= \mu_f$) and define $\mu_0=m_{W_H}$. We employ CTEQ6L1 and CTEQ6M
in the the LO and NLO calculations respectively \cite{cteq}, and fix the LHT
parameters $\kappa=1$ and $s_\alpha = c_\alpha =
\frac{\sqrt{2}}{2}$. In this work we only use the PROSPINO scheme
to deal with the real-quark emission subprocesses.

\par
\subsection{Dependence on factorization/renormalization scale  }
\par
\begin{figure}[htbp]
\begin{center}
\includegraphics[width=0.45\textwidth]{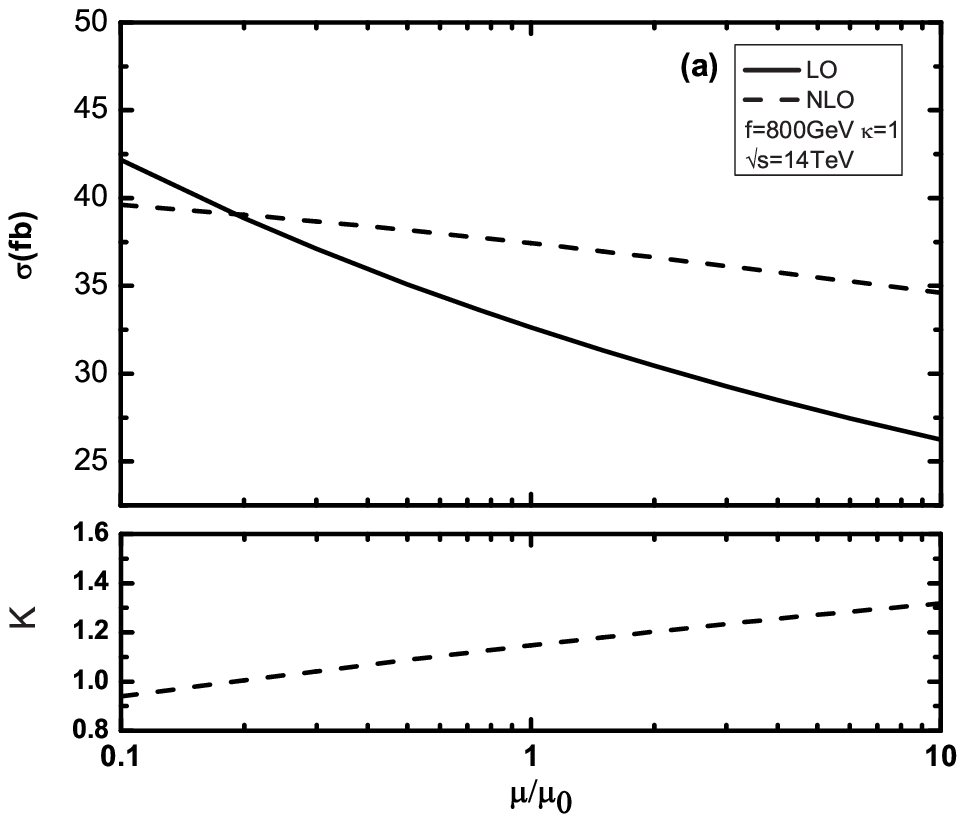}
\includegraphics[width=0.45\textwidth]{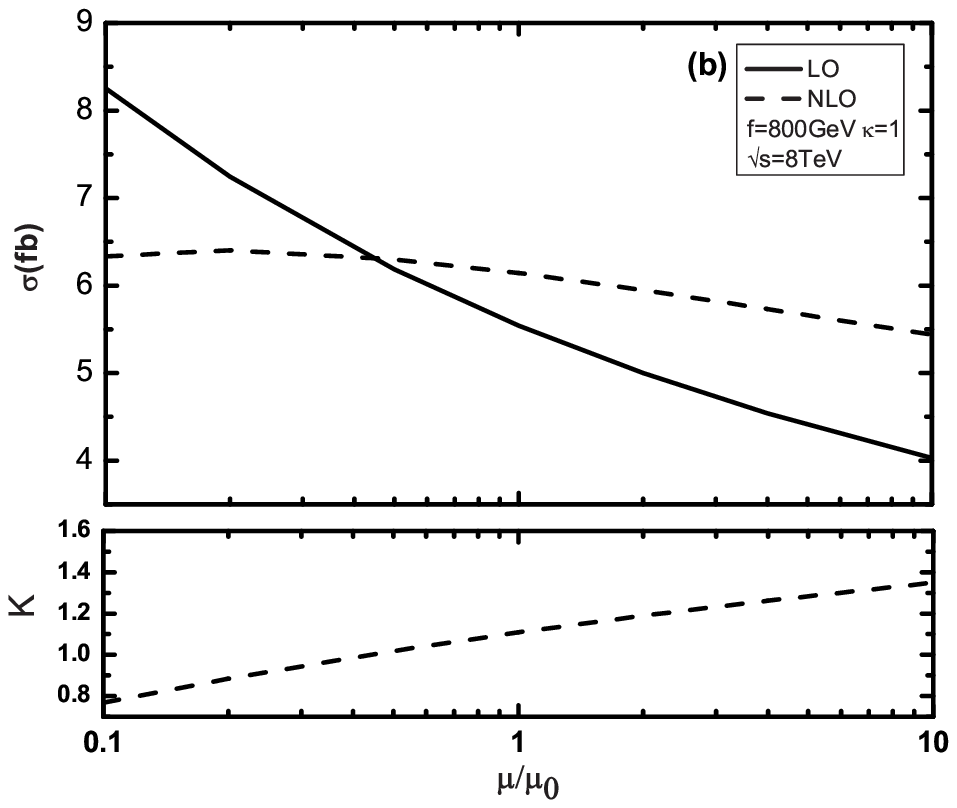}
\hspace{0in}%
\caption{\label{fig6} The dependence of the LO, QCD NLO corrected integrated
cross sections and the corresponding $K$-factors for the \ppww process on the
factorization/renormalization scale $\mu$ at the LHC with $f=800~GeV$,
$\kappa=1$ and $s_\alpha = c_\alpha = \frac{\sqrt{2}}{2}$.
(a) $\sqrt{s}=14~TeV$. (b) $\sqrt{s}=8~TeV$. }
\end{center}
\end{figure}

In Figs.\ref{fig6}(a,b) we present the dependence of the LO, QCD NLO
corrected integrated cross sections and the corresponding $K$-factor
($K\equiv \sigma_{NLO}/\sigma_{LO}$) on the
factorization/renormalization scale $\mu$ for the process \ppww at
the $\sqrt{s}=14~TeV$ and the $\sqrt{s}=8~TeV$ LHC separately, where
we take the LHT parameters $f=800~GeV$, $\kappa = 1$ and $s_\alpha =
c_\alpha = \frac{\sqrt{2}}{2}$. From the curves in
Figs.\ref{fig6}(a,b), we find that QCD NLO corrections to the $pp
\to W_H^+ W_H^- + X$ process significantly reduce the scale
uncertainty. We can read out from the figures that the LO and QCD
NLO corrected cross sections at $\mu_0=m_{W_H}$ are
$\sigma_{LO}(\sqrt{s}=14TeV)=32.63^{+9.56}_{-6.38} ~fb$,
$\sigma_{NLO}(\sqrt{s}=14TeV)=37.43^{+2.19}_{-2.83} ~fb$ and
$\sigma_{LO}(\sqrt{s}=8TeV)=5.54^{+2.71}_{-1.51} ~fb$,
$\sigma_{NLO}(\sqrt{s}=8TeV)=6.14^{+0.26}_{-0.70} ~fb$, where the
uncertainties describe the missing higher-order corrections
estimated via scale variations in the range of $0.1 \mu_0 < \mu <
10\mu_0$. The $K$-factor varies from $0.94~(0.77)$ to $1.32~(1.35)$
at the $\sqrt{s}=14~TeV$ ($8~TeV$) LHC, when $\mu/\mu_0$ goes from
$0.1$ to $10$. With the definition of scale uncertainty as
$\eta=\frac{|\sigma(0.1\mu_0)-\sigma(10\mu_0)|}{\sigma(\mu_0)}$, we
obtain that the scale uncertainties are reduced from $48.88\%$ (LO)
to $13.40\%$ (NLO) at the $\sqrt{s}=14~TeV$ and from $76.23\%$ (LO)
to $14.54\%$ (NLO) at the $\sqrt{s}=8~TeV$ LHC, respectively. In
Table \ref{tab1-1} we list some numerical results of the cross
sections and $K$-factors for some typical values of $\mu/\mu_0$,
which are read out from Figs.\ref{fig6}(a,b). In order to
investigate the contribution from the $pp \to gg \to W_H^+W_H^-+X$
process, which is considered as a component of the QCD NLO
corrections to the parent process \ppww, we also present the cross
sections for the $pp \to gg \to W_H^+W_H^-+X$ process ($\sigma(gg)$)
in this table. We can obtain from the data that the QCD NLO
correction part from the $pp \to gg \to W_H^+W_H^-+X$ process at
$\mu=\mu_0$ is about $5.85\%$ ($3.23\%$) of the total QCD NLO
correction ($\Delta \sigma_{NLO}$) at the $14~TeV~(8~TeV)$ LHC. We
can see also that the NLO theoretical uncertainty due the choice of
$\mu$ mainly comes from the genuine QCD NLO corrected cross section
for the $pp \to q\bar{q} \to W_H^+W_H^-+X$ process, while the
contribution from the $pp \to gg \to W_H^+W_H^-+X$ process is
relatively small. In further numerical calculations we fix the
renormalization and factorization scales being equal to their
central value, i.e., $\mu=\mu_r=\mu_f=\mu_0=m_{W_H}$.
\begin{table}
\begin{center}
\begin{tabular}{c|c|c|c|c|c}
\hline
$\sqrt{s}$&$\mu/\mu_0$&$\sigma_{LO}$&$\sigma_{NLO}$&$\sigma(gg)$&$K$\\
$(TeV)$& &$(fb)$&$(fb)$&$(fb)$&\\
\hline
 &0.1&42.190(1)&39.62(2)&0.993(1)&0.939\\
 &0.5&35.091(1)&38.17(2)&0.3946(6)&1.09\\
14&1&32.626(1)&37.43(2)&0.2810(5)&1.15\\
 &2&30.444(1)&37.53(2)&0.2056(4)&1.20\\
 &10&26.242(1)&34.60(2)&0.1081(2)&1.32\\
\hline
 &0.1&8.2548(4)&6.333(3)&0.0805(1)&0.767\\
 &0.5&6.1850(3)&6.300(3)&0.02842(3)&1.019\\
8&1&5.5417(2)&6.143(3)&0.01943(3)&1.11\\
 &2&5.0006(2)&5.947(3)&0.01371(2)&1.21\\
 &10&4.0304(2)&5.440(2)&0.00671(1)&1.40\\
\hline
\end{tabular}
\end{center}
\begin{center}
\begin{minipage}{15cm}
\caption{\label{tab1-1} The numerical results of $\sigma_{LO}$,
$\sigma_{NLO}$ and the corresponding $K$-factors at the
$14~TeV$ and the $8~TeV$ LHC by taking $f=800~GeV$, $\kappa =
1$, $s_\alpha = c_\alpha = \frac{\sqrt{2}}{2}$ and some typical
values of factorization/renormalization scale $\mu$.
$\sigma(gg)$ is the cross section for the
$pp \to gg \to W_H^+W_H^-+X$ process, which is considered as a component
of the QCD NLO correction to the parent process \ppww. }
\end{minipage}
\end{center}
\end{table}

\par
\subsection{Dependence on global symmetry breaking scale $f$ }
\par
\par
\begin{figure}[htbp]
\begin{center}
\includegraphics[width=0.45\textwidth]{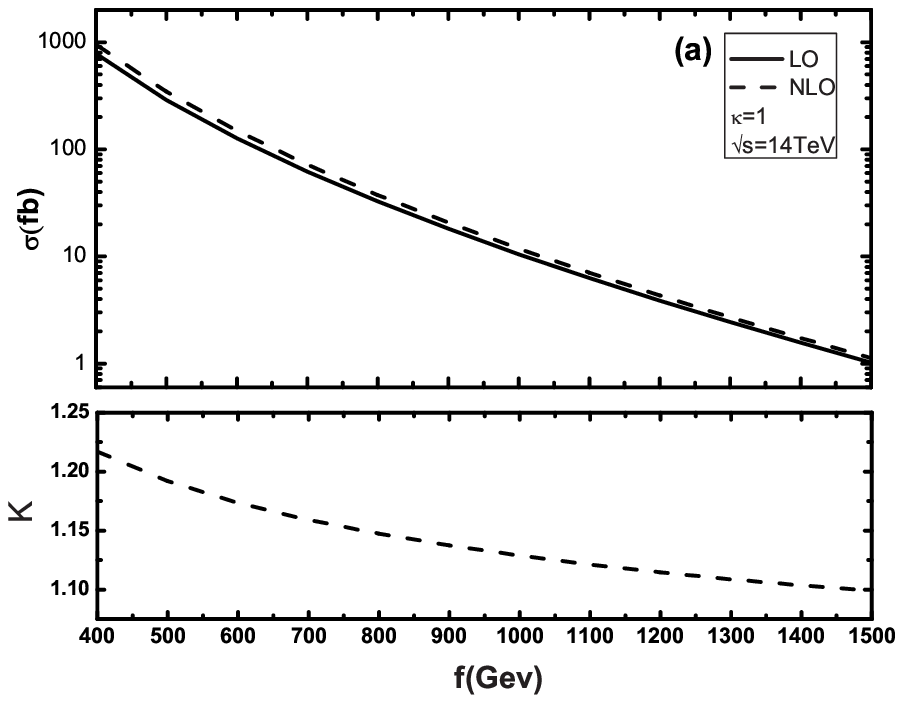}
\includegraphics[width=0.45\textwidth]{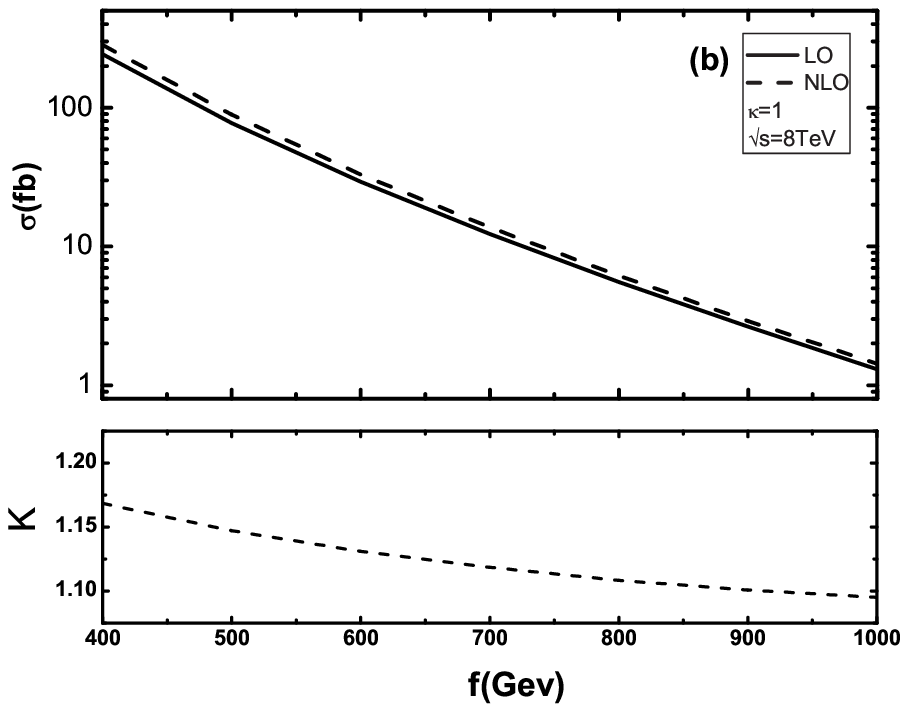}
\hspace{0in}%
\caption{\label{fig7} The LO, QCD NLO corrected integrated cross sections
and the corresponding $K$-factors for the \ppww process as the functions
of the global symmetry breaking scale $f$ at the LHC with $\kappa =1$ and
$s_\alpha = c_\alpha = \frac{\sqrt{2}}{2}$. (a) $\sqrt{s}=14~TeV$.
(b) $\sqrt{s}=8~TeV$. }
\end{center}
\end{figure}

The LO and QCD NLO corrected integrated cross sections together with
the corresponding $K$-factor as functions of the scale $f$ at the
$\sqrt{s}=14~TeV$ and the $\sqrt{s}=8~TeV$ LHC are depicted in
Figs.\ref{fig7}(a) and (b), respectively. We can see from
Fig.\ref{fig7} that the LO and NLO total cross sections for the
\ppww process decrease drastically when $f$ goes up. This is because
the mass of final $W_H$ becomes heavier as the increment of $f$,
therefore the phase-space becomes smaller. We can read out from the
figures that the corresponding $K$-factor varies from $1.22$ to
$1.10$ at the $\sqrt{s}=14~TeV$ LHC and from $1.17$ to $1.10$ at the
$\sqrt{s}=8~TeV$ LHC in the plotted $f$ range. In Table \ref{tab3}, we
list some numerical results of the LO, NLO cross sections and the
corresponding $K$-factors for some typical values of $f$ which are
shown in Figs.\ref{fig7}(a,b).
\begin{table}
\begin{center}
\begin{tabular}{c|c|c|c|c}
\hline
$\sqrt{s}$  & $f$ & $\sigma_{LO}$  & $\sigma_{NLO}$ & $K$  \\
$(TeV)$ & $(GeV)$ & $(fb)$ & $(fb)$ &  \\
\hline
     & 500  & 289.93(1)    &345.6(2)      &1.19    \\
     & 700  & 62.053(3)    &71.93(4)      &1.16    \\
     & 800  & 32.626(1)    &37.43(2)      &1.15    \\
 14  & 900  & 18.1252(8)   &20.61(1)      &1.14    \\
     & 1100 & 6.2964(3)    &7.059(3)      &1.12    \\
     & 1300 & 2.44312(9)   &2.709(1)      &1.11    \\
     & 1500 & 1.02314(4)   &1.1246(5)     &1.10    \\
\hline
    & 500  & 77.693(3)     &89.12(5)      &1.15    \\
  8 & 700  & 12.2863(5)    &13.741(7)     &1.12    \\
    & 800  & 5.5417(2)     &6.143(3)      &1.11    \\
    & 900  & 2.6357(1)     &2.901(1)      &1.10    \\
\hline
\end{tabular}
\end{center}
\begin{center}
\begin{minipage}{15cm}
\caption{\label{tab3} The numerical results of $\sigma_{LO}$,
$\sigma_{NLO}$ and the corresponding $K$-factors at the
$\sqrt{s}=14~TeV$ and the $\sqrt{s}=8~TeV$ LHC by taking $\kappa =
1$, $s_\alpha = c_\alpha = \frac{\sqrt{2}}{2}$, $\mu=\mu_0$ and some typical
values of $f$.  }
\end{minipage}
\end{center}
\end{table}

\vskip 5mm
\section{Summary}
\par
In this paper, we present the calculation of $W_H(Z_H) q_-$ and $W_H$-pair
productions at the $\sqrt{s}=14~TeV$ and the $\sqrt{s}=8~TeV$ LHC in
QCD NLO. The dependence of the total cross section on the
renormalization/factorization scale shows that the QCD NLO
corrections can reduce significantly the uncertainty of the LO
theoretical predictions. We also display the dependence of cross sections
on global symmetry breaking scale $f$.
The more calculation detail and numerical results were presented
in Refs.\cite{YanH,DuSM}.

\vskip 5mm
\par
\noindent{\large\bf Acknowledgments:} This work was supported in
part by the National Natural Science Foundation of China (Contract
No.11075150, No.11005101), and the Specialized Research Fund for the
Doctoral Program of Higher Education (Contract No.20093402110030).


\vskip 5mm
\begin{thebibliography}{99}
\bibitem{s1}
  S. L. Glashow, Nucl. Phys. {\bf 22} (1961) 579;
  S. Weinberg, Phys. Rev. Lett. {\bf 19} (1967) 1264;
  A. Salam, Proc. 8th Nobel Symposium Stockholm 1968,ed. N. Svartholm
  (Almquist and Wiksells, Stockholm 1968) p.367;
  H. D. Politzer, Phys. Rept. {\bf 14} (1974) 129.

\bibitem{s2}
  P. W. Higgs, Phys. Lett. {\bf 12} (1964) 132, Phys. Rev. Lett. {\bf 13} (1964)
  508, Phys. Rev. {\bf 145} (1966) 1156;
  F. Englert and R. Brout, Phys. Rev. Lett. {\bf 13} (1964) 321;
  G. S. Guralnik, C. R. Hagen and T. W. B. Kibble, Phys. Rev. Lett. {\bf 13} (1964) 585;
  T. W. B. Kibble, Phys. Rev. {\bf 155} (1967) 1554.

\bibitem{model-1}
  G.G. Ross, ¡°Grand Unied Theories¡° (Addison-Wesley Publishing
  Company, Reading, MA, (1984); P. Langacker, Phys. Rep. 72 (1981)
  185; H. Georgi, S.L Glashow, Phys. Rev. Lett. 32 (1974) 438; A.J.
  Buras, J. Ellis, M.K. Gaillard, D.V. Nanopoulos, Nucl. Phys. {\bf
  B135}(1978) 66.

\bibitem{model-2}
  S. P. Martin, arXiv:hep-ph/9709356; M. E. Peskin,
  arXiv:0801.1928 [hep-ph]; K. A. Olive, arXiv:hep-ph/9911307; M.
  Drees, arXiv:hep-ph/9611409; H. E. Haber and G. L. Kane, Phys. Rept.
  117 (1985) 75; H. P. Nilles, Phys. Rept. 110, 1 (1984); A. Signer,
  J.Phys.{\bf G36} (2009) 073002, arXiv:0905.4630 [hep-ph].

\bibitem{model-3}
  N. Arkani-Hamed, S. Dimopoulos, G. Dvali, Phys. Lett. B429 (34):
  263(1998), arXiv:hep-ph/9803315; N. Arkani-Hamed, S. Dimopoulos,
  G. Dvali, Phys. Rev. D59 086004(1999), arXiv:hep-ph/9807344; I.
  Antoniadis, N. Arkani-Hamed, S. Dimopoulos, G. Dvali, Phys. Lett.
  {\bf B436} (34): 257(1998), arXiv:hep-ph/9804398; M. Shifman,
  Int. J. Mod. Phys. {\bf A25} 199-225,2010, arXiv:0907.3074v2 [hep-ph].

\bibitem{model-5}
  R. N. Mohapatra and J. C. Pati, Phys. Rev. {\bf D11} (1975), 566571;
  G. Senjanovic and R. N. Mohapatra; Phys. Rev. {\bf D12} (1975),
  1502; A. Adulpravitchai, M. Lindner, A. Merle, and R. N. Mohapatra,
  Phys. Lett. {\bf B680} (2009) 476, arXiv/hep-ph:0908.0470.

\bibitem{model-6}
  L.Basso, S. Moretti, G.M. Pruna, Phys. Rev.{\bf D83}(2011) 055014,
  arXiv:1011.2612v4 [hep-ph].

\bibitem{LittleHiggs}
  N. Arkani-Hamed, A. G. Cohen and H. Georgi, Phys. Lett. {\bf B513} (2001) 232;
  M. Schmaltz and D. Tucker-Smith, Ann. Rev. Nucl. Part. Sci. {\bf 55} (2005) 229;
  M. Perelstein, Prog. Part. Nucl. Phys. {\bf 58} (2007) 247; and references therein.

\bibitem{Arkani}
  Arkani-Hamed, A. G. Cohen, E. Katz, A. E. Nelson, T. Gregoire and
  J. G. Wacker, JHEP {\bf 08} (2002) 021.

\bibitem{LH2}
  N. Arkani-Hamed, A. G. Cohen, T. Gregoire, J. G. Wacker and A. G. Cohen,
  JHEP {\bf 08} (2002) 020.

\bibitem{LH5}
  I. Low, W. Skiba and D.Smith, Phys. Rev. {\bf D66},
  (2002) 072001.

\bibitem{Limit-f}
  C. Csaki, J. Hubisz, G. D. Kribs, P. Meade and J. Terning, Phys. Rev. {\bf D67}
  (2003) 115002, [arXiv:hep-ph/0211124].

\bibitem{LH6}
  M. Schmaltz, Nucl. Phys. Proc. Suppl. {\bf 117} (2003) 40.

\bibitem{LH7}
  T. Gregoire and J. G. Wacker, JHEP {\bf 08} (2002) 019.

\bibitem{LH4}
  N. Arkani-Hamed, A. G. Cohen, E. Katz and A. E. Nelson,
  JHEP {\bf 07} (2002) 034.

\bibitem{Wmass}
 ATLAS Collaboration, Phys. Lett. {\bf B705} (2011) 18-46.

\bibitem{Zmass}
 D. Olivito, for the ATLAS collaboration, Conference proceedings for the Meeting
 of the Division of Particles and Fields of the American Physical Society (DPF), 2011, arXiv:1109.0934.

\bibitem{Low:2004xc}
  I. Low, JHEP {\bf 0410} (2004) 067.

\bibitem{Barbieri:2000gf}
  R. Barbieri and A. Strumia, IFUP-TH/2000-22 and SNS-PH/00-12, [arXiv:hep-ph/0007265].

\bibitem{Hubisz:2004ft}
  J. Hubisz and P. Meade, Phys. Rev. {\bf D71} (2005) 035016.

\bibitem{Hubisz:2005tx}
  J. Hubisz, P. Meade, A. Noble and M. Perelstein, JHEP {\bf 01} (2006) 135.

\bibitem{Cheng:2003ju}
  H.-C. Cheng and I. Low, {\bf JHEP} 09 (2003) 051; {\bf JHEP} 08 (2004) 061.

\bibitem{Asano}
  A. Birkedal, A. Noble, M. Perelstein and A. Spray, Phys. Rev. {\bf D74} (2006) 035002;
  M. Asano, S. Matsumoto, N. Okada, and Y. Okada,  Phys. Rev. {\bf D75} (2007) 063506.

\bibitem{cpyuan:2006ph}
  A. Belyaev, C. -R. Chen, K. Tobe and C. -P. Yuan,
  Phys. Rev. {\bf D74}, (2006) 115020.


\bibitem{sasha_pheno}
  A. Belyaev, C. -R. Chen, K. Tobe and C. -P. Yuan, in Proceedings of Monte Carlo
  Tools for Beyond the Standard Model Physics, Fermilab, 2006,
  given by A. Belyaev, http://theory.fnal.gov/mc4bsm/agenda.html; in Proceedings of
  Osaka University, 2006, Osaka, given by C. -P. Yuan,
  http://www-het.phys.sci.osaka-u.ac.jp/seminar/seminar/seminar.html;
  in Proceedings of the Summer Institute on Collider Phenomenology,
  National Tsing Hua University, Taiwan, 2006, given by K. Tobe,
  http://charm.phys.nthu.edu.tw/hep/summer2006/;
  in Proceedings of ICHEP'06, Moscow, 2006, given by A. Belyaev,
  http://ichep06.jinr.ru/reports/116$\_$11s1$\_$10p20$\_$belyaev.pdf.

\bibitem{Chen:2006ie}
  C.-S. Chen, K. Cheung and T. -C. Yuan, Phys. Lett. {\bf B644} (2007) 158.

\bibitem{YanH}
  R.-Y. Zhang, H. Yan, W.-G. Ma, S.-M. Wang, L. Guo and L. Han,
  Phys. Rev. {\bf D85} (2012)015017.

\bibitem{DuSM}
  S.-M. Du, L. Guo, W. Liu, W.-G. Ma and R.-Y. Zhang,
  Phys. Rev. {\bf D86} (2013) 054027.


\bibitem{Hubisz}
  J. Hubisz, P. Meade, A. Noble and M. Perelstein, JHEP {\bf 0601}, 135
  (2006), arXiv:hep-ph/0506042.





\bibitem{Blanke:2007ckm}
  M. Blanke, A. J. Buras, A. Poschenrieder, S. Recksiegel, C.
  Tarantino, S. Uhlig and A. Weiler JHEP {\bf 01} (2007) 066.

\bibitem{gs}
  J. Collins, F. Wilczek, and A. Zee, Phys. Rev. {\bf D18} (1978) 242;
  W. J. Marciano, Phys. Rev. {\bf D29} (1984) 580;
  P. Nason, S. Dawson and R.K. Ellis, Nucl. Phys. {\bf B327} (1989) 49,
  Nucl. Phys. {\bf B335} (1990) 260(E).

\bibitem{PROSPINO-ref}
  W. Beenakker, R. H\"opker, M. Spira and P. M. Zerwas, Nucl. Phys. {\bf B492}
  (1997) 51; W. Beenakker, M. Klasen, M. Kr\"amer, T. Plehn, M. Spira and P. M. Zerwas, Phys.
  Rev. Lett. {\bf 83} (1999) 3780;
  http://www.thphys.uni-heidelberg.de/\~~plehn/index.php?show=prospino.

\bibitem{on-shell subtraction}
   T. Plehn and C. Weydert, PoS {\bf CHARGED2010} (2010) 026, [arXiv:1012.3761];
   T. Binoth, D. Goncalves-Netto, D. Lopez-Val, K. Mawatari, T. Plehn and I.
   Wigmore, Phys. Rev. D 84 (2011) 075005, [arXiv:1108.1250].

\bibitem{databook}
  K. Nakamura, {\it et al.}, J. Phys. {\bf G37} (2010) 075021.

\bibitem{cteq}
  J. Pumplin, D. R. Stump, J. Huston, H. -L. Lai, P. Nadolsky and W. -K.
  Tung, JHEP {\bf 07} (2002) 012;
  D. Stump, J. Huston, J. Pumplin, W. -K. Tung, H. -L. Lai, S. Kuhlmann and J. F.
  Owens, JHEP {\bf 10} (2003) 046.

\end{thebibliography}
\end{document}